\documentclass[12pt]{article}
\usepackage{times}

\begin{document}

\def\bA{{\bf A}}
\def\bP{{\bf P}}
\def\bE{{\bf E}}
\def\ba{{\bf a}}
\def\be{{\bf e}}
\def\bv{{\bf v}}
\def\bk{{\bf k}}

\bibliographystyle{prsty}

\title{Improvements of the Discrete Dipole Approximation method}

\author{Piotr J. Flatau \\
Scripps Institution of Oceanography, University of California, San
Diego, \\La Jolla, California 92093-0221}

\maketitle

\begin{abstract}
We report improvements in complex conjugate gradient algorithms
applied to the discrete dipole approximation (DDA). It is shown
that computational time is reduced by using  the Bi-CGSTAB version
of the CG algorithm, with diagonal left preconditioning.


\begin{center}
Optics Letters 1997, volume 22, number 16, 1205-1207.
\newline {\copyright\ Optical Society of America, 1997.}
\end{center}
\end{abstract}

The discrete-dipole approximation (DDA) is a flexible technique
for computing scattering and absorption by targets of arbitrary
geometry. In \cite{Draine94a} the discrete dipole approximation
(DDA) for scattering calculations is reviewed. Rather than
``direct'' methods for solving linear system of equations arising
in DDA problem iterative methods for finding the solution have
proven effective and efficient.

In this paper we perform systematic study of various
non-stationary iterative (conjugate gradient) methods in search
for the most efficient one. We document implementation of these
methods in our public domain code DDSCAT.5a code\cite{Draine94a}

Numerical aspects of the discrete dipole approximation continue to
be of great interest. Yung \cite{Yung78a} applied a conjugate
gradient method to the in DDA approach. Hoekstra
\cite{Hoekstra94b} identifies Yung's scheme as the conjugate
gradient (CG) algorithm proposed by Hestenes  \cite{Hes52a}.
Rahola \cite{Rahola96a} discusses solution of dense systems of
linear equations in the discrete-dipole approximation and choice
of of the best iterative method in this application. Draine
\cite{Draine88a} implemented a conjugate gradient method based on
work of Petravic  and Kuo-Petravic. \cite{Petravic79a} This
implementation is quite robust and has been  used for many years.
\cite{Draine94a} However, Lumme and Rahola \cite{Lumme94a} applied
the quasi-minimal residual (QMR) conjugate gradient algorithm to
the system of linear equations arising in the DDA applications.
They claim that the QMR method is approximately 3 times faster in
comparison to the one  employed in the DDSCAT code.
\cite{Draine94a} It was this work which prompted us to perform the
analysis reported here.

PIM\cite{Cunha95a} is a collection of Fortran 77 routines designed
to solve systems of linear equations  on parallel and sequential
computers using a variety of iterative methods.

PIM contains implementations of various methods:
conjugate-gradient (CG); Conjugate-Gradients for normal equations
with minimization of the residual norm (CGNR); Conjugate-Gradients
for normal equations with minimization of the error norm (CGNE);
Bi-Conjugate-Gradients (Bi-CG); Conjugate-Gradients squared (CGS);
the stabilised version of Bi-Conjugate-Gradients (Bi-CGSTAB); the
restarted, stabilised version of Bi-Conjugate-Gradients
(RBi-CGSTAB); the restarted, generalized minimal residual
(RGMRES); the restarted, generalized conjugate residual (RGCR),
the quasi-minimal residual with coupled two-term recurrences
(QMR); the transpose-free quasi-minimal residual (TFQMR); and
Chebyshev acceleration. The routines allow the use of
preconditioners; the user may choose to use left-, right- or
symmetric-preconditioning.

The convergence rate of iterative methods depends on the
coefficient matrix. Hence one may attempt to transform the linear
system into one that is equivalent (in the sense that it has the
same solution) but is easier to solve. A preconditioner is a
matrix $M$ that effects such a transformation. It is possible to
introduce left- and right preconditioners.\cite{Barrett94a} The
simplest preconditioner consists of just the diagonal of the
coefficient matrix. This is known as the (point) Jacobi
preconditioner.

To compare these different algorithms we have used them to find
solutions to the problem of scattering by a homogeneous sphere.
The scattering problem is specified by the usual size parameter
$x= 2 \pi a / \lambda$, where a is the radius.

Tables~\ref{table1} and \ref{table4} presents the number of
iterations and CPU time for size parameter $x=0.1$ and $x=1$ and
for several values of refractive index. The conjugate gradient
methods are defined as above. Label (L) indicates left Jacobi
preconditioning. For example CGNE(L) is the conjugate gradient
method  for normal equations with minimization of the error norm
and left Jacobi preconditioning. Similarly, (R) indicates right
Jacobi preconditioning. CPU time (sequential Silicon Graphics
workstation) is normalized to the ``best'' method. Star indicates
that the method did not converge in the maximum allowed number of
iterations or that the method failed to converge. Fractional error
$10^{-5}$ was used as the stopping criterion. The DDSCAT.5a
code\cite{Draine94a}  with the newly implemented GPFA fast Fourier
transform method was used. For Bi-CGSTAB and CGNE we used left and
right Neumann polynomial preconditioner truncated after the first
term. Thus, Bi-CGSTAB(N)(L)  indicates the stabilised version of
Bi-Conjugate-Gradients method with left Neumann polynomial
preconditioner.

Table~\ref{table1} presents results for size parameter $x=0.1$ and
real refractive index ${\rm n}=1.33, 2, 3, 5$ as well as one case
with small complex part of refractive index ${\rm n}=(5, 0.0001)$
and  size parameter ${\rm x}=0.1$. In Table 1  the  CPU times  are
normalized to the CG(L)  method, which was found to be the best
method. For example it is 4.0 times faster in comparison with the
CGNE for ${\rm n}=(1.33,0)$. For larger values of real refractive
index the CGNE is almost an order of magnitude slower in
comparison to CG. This is because more iterations are needed for
the same convergence and because cost of one CG iteration is less
than cost of one CGNE iteration. The QMR algorithm is never
competitive and actually fails to converge for large real
refractive indices. For small refractive index the Bi-CGSTAB
algorithm is comparable to the CG and requires less iterations.
However, the cost per iteration is larger in comparison to CG
which offsets the advantage of lesser number of iterations. The
Petravic  and Kuo-Petravic \cite{Petravic79a} algorithm used by us
for many years \cite{Draine88a} is similar to CGNR and CGNE.
However, we observed on occasion slightly different convergence
rates due to stabilization of Petravic  and Kuo-Petravic algorithm
every 10th time step. \cite{Draine88a} This is true for all other
cases. The storage requirements of CG,  CGNE, CGNR is $6 \times
N$, for BiCG it is $8 \times N$, for CGS, Bi-CGSTAB, TFQMR it is
$10 \times N$, QMR requires $11 \times N$. Thus, for pure real
refractive index, the  CG is not only the fastest method but also
it requires the least amount of temporary storage. It can be seen
that left preconditioning by the inverse of diagonal of the DDA
matrix \cite{Draine94a}  reduces the number of iterations needed.
The added time needed for division by diagonal elements is
generally negligible in comparison to the time saved by smaller
amount of iterations. It can be seen that for  Bi-CGSTAB,
Bi-CGSTAB(L), and Bi-CGSTAB(R) the  left Jacobi preconditioning is
the only method converging for larger refractive index. Restarted
methods (RBi-CBSTAB and RGCR) appear to be not competitive but
further study may be needed (we used the orthogonal base of 10
vectors for all restarted methods). The CG method is also
competitive in cases with small absorption (see last column of
Table~\ref{table1}). We have also calculated (not presented)
results for size parameter of $x=0.1$ and increasing complex part
of refractive index $n=(1.33,0), (1.33, 0.01), (1.33,0.1),
(1.33,1), (1.33, 2), (1.33,3)$. The BiCGSTAB(L), which proved to
be   the most robust method. However the CGS(L) is competitive and
faster for $n=(1.33, 3)$. Both CGS(L) and BiCGSTAB(L) require the
same amount of iteration for convergence and their cost is
similar. These methods are between 2.9 and 1.6 times faster in
comparison to CGNR --- the method used in DDSCAT code. The QMR and
TFQMR which Lumme and Rahola \cite{Lumme94a} claim to be faster in
comparison to CGNR and the DDSCAT implementations do not converge
on occasion and when they work they are only slightly better in
this case. As before, left Jacobi preconditioning is almost always
beneficial. The CG(L) algorithm is faster than BiCGSTAB(L) for
refractive index $n=(1.33,0), (1.33,0.01), (1.33,0.1)$.

Table~\ref{table4} is for size parameter $x=1$. All the results
are normalized to Bi-CGSTAB(L). This method is clearly superior to
the CGNR method and it  is 2-4.3 faster. It can be seen that CGNR
converges slowly, and has not satisfied the stopping criterion in
140 iterations for $n=(3,0.0001)$. For this larger value of size
parameter the QMR algorithm doesn't converge well but its smooth
version TFQMR does. However, TFQMR is slower in comparison to
Bi-CGSTAB(L) and comparable to CGNR. The CG(L) method for
refractive index $n=(1.33,0)$ and $n=1.33,0.01$ is faster than the
reference scheme Bi-CGSTAB(L). It can be seen that the Neumann
polynomial preconditioning Bi-CGSTAB(N)(L) or Bi-CGSTAB(N)(R) does
reduce the number of iterations needed for certain cases of
refractive index.  However the cost associated with the additional
calculations always offsets this improved convergence rate. As
before, the  left Jacobi preconditioner is superior to right or
no-preconditioner cases. CG(L) works well for small refractive
index but is comparable to Bi-CGSTAB(L). The QMR algorithm fails
to converge but the transpose-free quasi-minimal residual (TFQMR)
algorithm converges well and is comparable to CGNR. The CG method
is theoretically valid for Hermitian positive definite matrices.
The matrix arising in the DDA is not Hermitian but symmetric.
Therefore, strictly speaking, the CG method  is not valid for use
in the DDA. The users are advised to test the CG  method when
extrapolating results presented here to different size parameters,
particle sizes,  and  refractive index values.
\begin{table}[ht]
\caption{\label{table1}CPU time (normalized)  and number of
iterations for x=0.1.}
\begin{tabular}{lccccc}
 Method  & n=(1.33,0) & (2,0) & (3,0) & (5,0) & (5,0.0001)  \\
\hline CGNE &    4.0(9) &    4.9(24) &    8.7(76) &  *(540) &
*(540) \\ CGNE(L) &    3.3(7) &    4.0(19) &    7.8(67) &  *(540)
&  *(540) \\ CGNE(R) &    4.1(9) &    4.9(24) &    8.8(76) &
*(540) &  *(540) \\ CGNE(N)(L) &    3.1(4) &    4.1(13) &
19.3(113) &  *(540) &  *(540) \\ CGNE(N)(R) &    4.4(6) & 7.7(25)
&  *(140) &  *(540) &  *(540) \\ CGNR &    4.0(9) & 4.7(23) &
8.0(69) &  *(540) &  *(540) \\ CGNR(L) &    3.3(7) & 4.0(19) &
5.9(50) &    4.6(329) &    3.5(330) \\ CGNR(R) & 4.1(9) &
4.7(23) &    8.0(69) &  *(540) &  *(540) \\ QMR & 3.7(6) &
3.3(11) &    3.3(19) &  *(111) &  *(78) \\ QMR(L) & 2.6(4) &
2.8(9) &    2.7(15) &  *(75) &  *(92) \\ QMR(R) & 3.8(6) &
3.4(11) &    3.4(19) &  *(268) &  *(540) \\ CG & 1.4(6) &
1.2(11) &    1.2(20) &    1.2(163) &    1.1(213) \\ CG(L) &
1.0(4) &    1.0(9) &    1.0(16) &    1.0(138) & 1.0(182) \\ CG(R)
&    1.4(6) &    1.2(11) &    1.2(20) & 1.1(157) &    1.2(213) \\
BiCG &    2.3(6) &    2.1(11) &  *(140) &  *(540) &  *(540) \\
BiCG(L) &    1.6(4) &    1.8(9) &  *(140) & *(540) &  *(540) \\
BiCG(R) &    2.4(6) &    2.2(11) &  *(140) & *(540) &  *(540) \\
Bi-CGSTAB &    1.8(4) &    1.5(7) &    1.5(13) &  *(540) &  *(540)
\\ Bi-CGSTAB(L) &    1.4(3) &    1.3(6) & 1.3(11) &    4.0(281) &
4.2(388) \\ Bi-CGSTAB(R) &    1.8(4) & 1.5(7) &    1.6(13) &
*(540) &  *(540) \\ Bi-CGSTAB(N)(L) & 1.9(2) &    6.8(17) &
14.9(65) &  *(540) &  *(540) \\ Bi-CGSTAB(N)(R) &    2.1(2) &
13.1(33) &   27.9(122) &  *(540) & *(540) \\ TFQMR &    3.8(5) &
3.4(9) &    4.3(19) &  *(540) & *(540) \\ TFQMR(L) &    3.1(4) &
3.1(8) &    4.1(18) &  *(540) &  *(540) \\ TFQMR(R) &    3.9(5) &
3.5(9) &    4.4(19) & *(540) &  *(540) \\ CGS &    1.7(4) &
1.5(7) &    1.4(12) & *(540) &  *(540) \\ CGS(L) &    1.4(3) &
1.3(6) &    1.2(10) & *(540) &  *(540) \\ CGS(R) &    1.8(4) &
1.5(7) &    1.4(12) & *(540) &  *(540) \\ RGCR &    4.3(2) &
2.8(2) &    2.5(3) & *(14) &  *(14) \\ RGCR(L) &    2.0(1) &
2.4(2) &    2.1(2) & *(14) &  *(14) \\ RGCR(R) &    4.4(2) &
2.8(2) &    2.7(3) & *(14) &  *(14) \\ RBi-CGSTAB &  *(12) &
*(12) &  *(12) &  *(12) & *(12) \\ RBi-CGSTAB(L) &  *(12) &  *(12)
&  *(12) &  *(12) & *(12) \\ RBi-CGSTAB(R) &  *(12) &  *(12) &
*(12) &  *(12) & *(12) \\ \hline
\end{tabular}
\end{table}
\begin{table}[ht]
\caption{\label{table4}CPU time (normalized)  and number of
iterations for x=1}
\begin{tabular}{lccccc}
 Method  & n=(1.33,0) & (1.33,0.01) & (1.33,1) & (2,0) & (3,0.0001)  \\
\hline CGNE &    3.2(10) &    3.2(10) &    2.0(16) &    4.5(33) &
*(140) \\ CGNE(L) &    2.7(8) &    2.6(8) &    1.7(13) & 3.8(27) &
*(140) \\ CGNE(R) &    3.2(10) &    3.2(10) & 2.0(16) &    4.6(33)
&  *(140) \\ CGNE(N)(L) &    2.6(5) & 2.6(5) &  *(140) &  *(140) &
*(140) \\ CGNE(N)(R) &    3.6(7) & 3.6(7) &  *(140) &  *(140) &
*(140) \\ CGNR &    3.5(11) & 3.5(11) &    2.0(16) &    4.3(32) &
*(140) \\ CGNR(L) &    2.7(8) &    2.6(8) &    1.7(13) &
3.7(27) &  *(140) \\ CGNR(R) & 3.5(11) &    3.5(11) &    2.0(16) &
4.4(32) &  *(140) \\ QMR & *(47) &  *(58) &  *(25) &  *(76) &
*(50) \\ QMR(L) &    5.3(12) & *(59) &  *(21) &  *(71) &  *(39) \\
QMR(R) &  *(52) &  *(63) & *(22) &  *(70) &  *(37) \\ CG &
1.3(8) &    1.3(8) &  *(140) & *(140) &  *(140) \\ CG(L) &
0.9(5) &    0.9(5) &  *(140) & *(140) &  *(140) \\ CG(R) &
1.3(8) &    1.3(8) &  *(140) & *(140) &  *(140) \\ BiCG &  *(140)
&  *(140) &  *(140) &  *(140) & *(140) \\ BiCG(L) &  *(140) &
*(140) &  *(140) &  *(140) & *(140) \\ BiCG(R) &  *(140) &  *(140)
&  *(140) &  *(140) & *(140) \\ Bi-CGSTAB &    1.3(4) &    1.3(4)
&    1.2(10) & 1.2(9) &    1.1(24) \\ Bi-CGSTAB(L) &    1.0(3) &
1.0(3) & 1.0(8) &    1.0(7) &    1.0(21) \\ Bi-CGSTAB(R) &
1.3(4) & 1.3(4) &    1.2(10) &    1.3(9) &    1.1(24) \\
Bi-CGSTAB(N)(L) & 1.4(2) &    1.4(2) &    1.5(6) &    4.6(17) &
*(140) \\ Bi-CGSTAB(N)(R) &    1.5(2) &    1.5(2) &    1.6(6) &
*(140) & *(140) \\ TFQMR &    3.3(6) &    3.3(6) &    3.3(14) &
3.4(13) &    3.8(42) \\ TFQMR(L) &    2.8(5) &    2.8(5) &
3.0(13) & 3.0(11) &    3.7(40) \\ TFQMR(R) &    3.4(6) &    3.4(6)
& 3.3(14) &    3.4(13) &    3.9(42) \\ CGS &    1.3(4) &    1.3(4)
& 1.3(11) &    1.4(10) &    1.6(34) \\ CGS(L) &    1.0(3) & 1.0(3)
&    1.2(10) &    1.3(9) &    1.1(23) \\ CGS(R) &    1.3(4) &
1.3(4) &    1.3(11) &    1.4(10) &    1.6(33) \\ RGCR & 3.7(2) &
4.5(2) &  *(14) &    3.3(3) &  *(14) \\ RGCR(L) & 3.2(2) &
3.1(2) &    6.5(6) &    2.9(3) &  *(14) \\ RGCR(R) & 3.7(2) &
3.7(2) &    7.4(7) &    3.3(3) &  *(14) \\ \hline
\end{tabular}
\end{table}

We recommend use of the stabilized version of the Bi-conjugate
gradient algorithm with left Jacobi preconditioning
[Bi-CGSTAB(L)]. This algorithms requires 67\% greater storage than
the CGNR algorithm, but is typically 2-3 times faster.

The recent version of Discrete Dipole Approximation code DDSCAT5a
developed by Draine and Flatau contains improvements documented in
this paper. The code is available via anonymous ftp from the
\verb|ftp.astro.princeton.edu| site or from the Light Scattering
and Radiative Transfer Codes Library --- \verb|SCATTERLIB|
(\verb|http://atol.ucsd.edu/~pflatau|).

I have been supported in part by the Office of Naval Research
Young Investigator Program and in part by DuPont Corporate
Educational Assistance. I would like to thank Drs M. J. Wolff and
A. E. Ilin who  helped with computer tests. Bruce Draine checked
the manuscript. Dr. R. J.  Riegert of Du Pont if acknowledged for
his continuing interest in DDSCAT developments.

\bibliography{all,cg,fft,local}

\end{document}